\documentclass[reprint,superscriptaddress,amsmath,amssymb,aps,pra]{revtex4-1}

\usepackage{graphicx}
\usepackage{natbib}
\usepackage{hyperref}
\usepackage{cleveref}
\usepackage[caption=false,labelformat=empty]{subfig}
\usepackage{mathrsfs}

\hypersetup{colorlinks=true,citecolor=blue}

\newcommand{\micron}{{\mu\mathrm{m}}}
\newcommand{\Wcm}{\mathrm{Wcm}}

\newcommand{\rmd}{\mathrm{d}}
\newcommand{\fs}{\mathrm{fs}}
\newcommand{\eV}{\mathrm{eV}}
\newcommand{\MeV}{\mathrm{MeV}}
\newcommand{\abs}[1]{\left|#1\right|}

\newcommand{\lnTwo}{\ln(2)}
\newcommand{\Power}{\mathcal{P}}
\DeclareMathOperator\erf{erf}
\newcommand{\aux}{\mathcal{R}}
\newcommand{\chicrr}{\chi_{c,\mathrm{rr}}}

\begin{document}

\title{Scaling laws for positron production in laser--electron-beam collisions}

\author{T. G. Blackburn}
\email{tom.blackburn@chalmers.se}
\affiliation{Department of Physics, Chalmers University of Technology, SE-41296 Gothenburg, Sweden}
\author{A. Ilderton}
\affiliation{Centre for Mathematical Sciences, Plymouth University, PL4 8AA, UK}
\author{C. D. Murphy}
\affiliation{York Plasma Institute, Department of Physics, University of York, York, YO10 5DD, UK}
\author{M. Marklund}
\affiliation{Department of Physics, Chalmers University of Technology, SE-41296 Gothenburg, Sweden}

\date{\today}

\begin{abstract}
Showers of $\gamma$-rays and positrons are produced when a high-energy electron
beam collides with a super-intense laser pulse. We present scaling laws for the
electron beam energy loss, the $\gamma$-ray spectrum, and the positron yield
and energy that are valid in the non-linear, radiation-reaction--dominated regime.
As an application we demonstrate that by employing the collision of a $>$GeV
electron beam with a laser pulse of intensity~$>5\times10^{21}\,\Wcm^{-2}$,
today's high-intensity laser facilities are capable of producing $O(10^4)$
positrons per shot via light-by-light scattering.
\end{abstract}

\maketitle

\section{Introduction}

Electron-positron pair creation by the interaction of light with light
is one of the simplest processes in quantum electrodynamics (QED).
Thus far, however, the two-photon process has not been detected in
experiment~\cite{BreitWheeler,Pike,Ribeyre}, and the observation of
multiphoton pair creation could only be accomplished by colliding
the high-energy electron beam of the SLAC facility and an intense
laser pulse~\cite{Bula,Burke,Bamber}. It is expected that the next
generation of high-intensity lasers~\cite{ELI,Vulcan,XCELS,Apollon}
will create electromagnetic fields of sufficient magnitude such
that the nonlinear analogue of the Breit-Wheeler process becomes
dominant~\cite{Hu} (see also~\cite{Ilderton:2010wr,King:2013osa}).
As such fields are only otherwise found in extreme astrophysical
environments~\cite{Goldreich,Piran,Ruffini,Harding}, the prospect
of studying plasma dynamics under these conditions in the laboratory
has attracted considerable interest~\cite{MourouRMP,MarklundRMP,DiPiazzaRMP}.

High-power laser facilities are already used to generate positrons
via the Bethe-Heitler process, which converts bremsstrahlung photons
produced by the deceleration of electrons in high-$Z$ material.
The energetic electrons themselves are generated via
direct illumination of the solid target~\cite{ChenExpt,ChenExpt2,LiangExpt}
or by laser-wakefield acceleration~\cite{GahnExpt,SarriExpt,SarriExpt2}.
In the case that the target electromagnetic field should be provided
purely by light, \citet{BellKirk} proposed an advantageous setup of
electrons accelerated by counter-propagating, circularly-polarized lasers,
which is anticipated to create critical-density pair plasmas for laser
intensities $>7\times10^{23}\,\Wcm^{-2}$~\cite{Duclous,Nerush,Elkina,Jirka,Grismayer}.
High-energy positrons may also be generated by the irradiation of a
solid~\cite{Ridgers} or near-critical target~\cite{Brady,Zhu}
with a laser of similar intensity.

	\begin{figure}
	\includegraphics[width=0.8\linewidth]{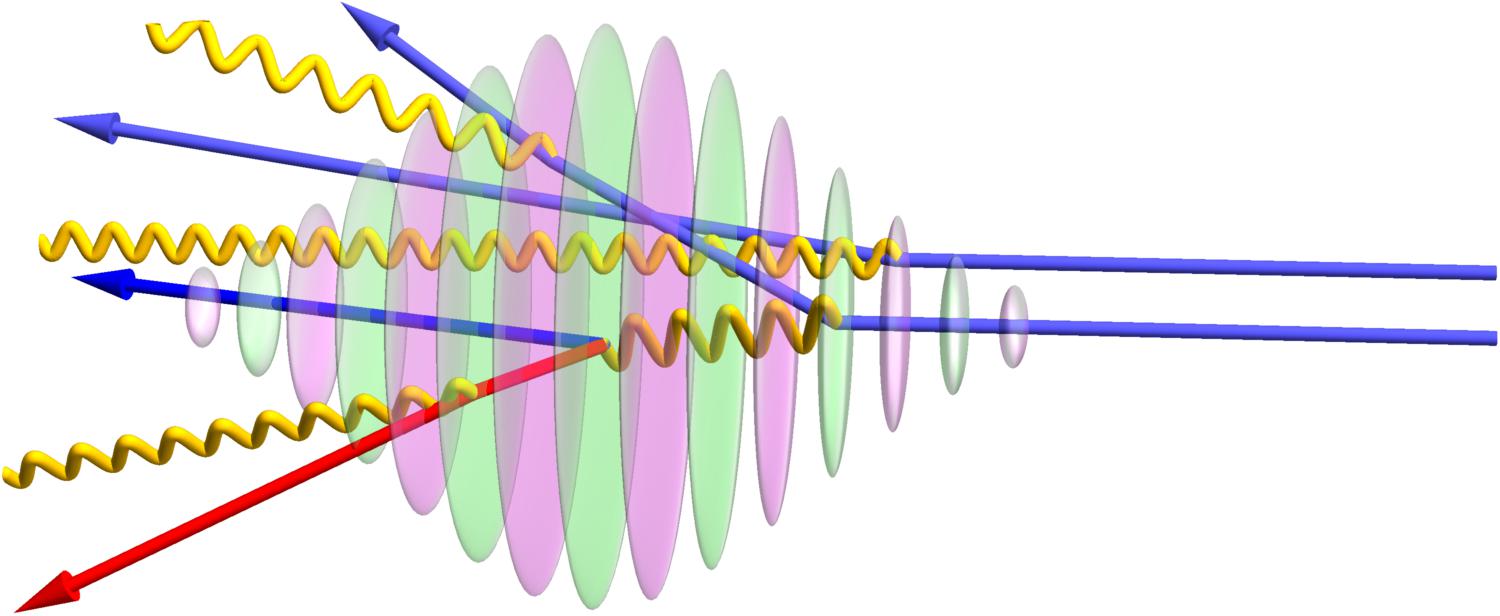}
	\caption[Particle shower]
		{Ultra-relativistic electrons (blue) collide with a
		counter-propagating laser pulse
		(magenta, green) and lose energy by emitting photons (yellow).
		Positrons (red) are created when photons undergo
		the non-linear Breit-Wheeler process.}
	\label{fig:shower}
	\end{figure}

Even though the highest intensity reached by currently available laser
systems ($10^{22}\,\Wcm^{-2}$~\cite{Hercules}) does not reach this level,
it is still possible to explore non-linear Breit-Wheeler pair creation
in these facilities by employing the head-on collision of high-energy
electron beam and intense laser pulse. This is because in the rest frame
of the electrons, the laser electric field amplitude is boosted by a
factor $\gamma \gg 1$. As the electron propagates through the laser pulse,
it loses energy by the emission of photons, which themselves interact
with the laser fields to produce pairs~\cite{Sokolov,Bulanov}, as shown
in \cref{fig:shower}. The experimental setup can be made entirely optical if
the necessary ultra-relativistic electron beam is obtained by
laser-wakefield acceleration~\cite{Wang,Kim,Leemans}. Such a configuration
has already been used to generate MeV $\gamma$-rays via non-linear
Thomson scattering~\cite{ChenS,SarriBeams} and has been studied as a
probe of quantum radiation reaction~\cite{Neitz,Blackburn,Dinu,Vranic}.

Here we consider the collision of a GeV electron beam with a laser pulse
of intensity $>10^{21}\,\Wcm^{-2}$ and present a set of analytical
scaling laws for the electron energy loss, the photon spectrum, and
the number and energy of positrons produced. This investigation will
complement the existing literature as it will bridge the gap between
analytic calculations in QED~\cite{Heinzl,King,Krajewska,Dinu:2013gaa,Meuren,Meuren2,DiPiazza,DiPiazza2}
and the use of large-scale particle-in-cell simulations~\cite{Mironov,Lobet}
that include QED processes by Monte Carlo sampling of rates
evaluated in the locally-constant-field approximation~\cite{RidgersJCP,Gonoskov}.

The paper is organized as follows. First we derive an approximation for
the pair creation probability of a single high-energy photon colliding
with an intense laser pulse in \cref{ssec:PhotonBeam}. Then we consider
producing these gamma rays via the inverse Compton scattering of an electron
beam. We show in \cref{sec:phi_c} that high-energy photon production is
maximized in the leading edge of the pulse near a point we call the
`effective centre''; identifying this region lets us estimate the
electron energy loss in \cref{sec:rr} and derive an expression for the
photon spectrum that accounts for radiation reaction in \cref{sec:PhotonSpectrum}.
We present scaling laws for the number and mean energy of the positrons
arising from pair creation of these photons in
\cref{ssec:ElectronBeam,ssec:MeanPositronEnergy} respectively. Finally, we
show how the colliding beams' finite sizes and offset affect the positron yield.

Natural units $\hbar = c = 1$ are used throughout.

\section{Pair creation probability for a single photon}
\label{ssec:PhotonBeam}

The importance of QED effects when photons and electrons interact with a
strong electromagnetic field is governed by the
quantum non-linearity parameter~\cite{Erber,Ritus,BKS}
	\begin{equation}
	\chi = \frac{e \sqrt{-(F\cdot p)^2}}{m^3}.
	\label{eq:ChiDef}
	\end{equation}
Here $e,m$ are the electron charge and mass, $F$ is the electromagnetic
field tensor and $p$ the particle four-momentum.
($\chi$ will be used to refer to electrons only; $\chi_\gamma$ and $\chi_+$
will be used for photons and positrons respectively.)
$\chi$ compares the magnitude of the electric field in the electron rest
frame to that of the critical field of QED
$E_\text{crit} = m^2/e = 1.326\times10^{18}\,\text{Vm}^{-1}$~\cite{Heisenberg}.
Even though the equivalent intensity, $3\times10^{29}\,\Wcm^{-2}$, is beyond
our present capability, it is possible to reach $\chi \sim 1$
by colliding ultra-relativistic particles with weaker fields.

We consider a linearly-polarized electromagnetic wave with Gaussian temporal
profile as a simple model of a laser pulse, with amplitude 
$E = (m \omega_0 a_0/e) \sin\phi\,\exp(-\lnTwo \phi^2/(2 \pi^2 n^2))$
at phase $\phi$.
Here $a_0$ is the usual dimensionless strength parameter,
$\omega_0 = 1.24~\eV/(\lambda/\micron)$ is the wave frequency,
and $n$ is the number of cycles corresponding to the pulse duration $\tau$,
defined to be the full width at half maximum (FWHM) of the intensity profile.
For now we consider only collisions with plane waves. We will introduce
a finite size for both the electron beam and laser pulse to reach
our final result in \cref{ssec:Expt}.

In electromagnetic fields with $a_0 \gg 1$, the formation length of QED
processes is much smaller than the characteristic length scale of the
external field, and we may use the locally-constant-field approximation
(LCFA)~\cite{Ritus}. This permits probabilities and rates to be calculated
in an equivalent system of constant fields that have the same local value
of $\chi$. It underlies numerical studies of highly-intense laser-plasma
interactions, where the electromagnetic fields have complex spatial and
temporal structure that make direct analysis from strong-field QED
unfeasible~\cite{RidgersJCP,Gonoskov}. The field structure we consider
here is much simpler, so the LCFA is key to keeping all our results analytical.
%Using the LCFA neglects interference phenomena that are important at
%$a_0 \sim 1$ and in the low-energy part of the photon spectrum~\cite{HarveyPRA,Dinu}.

When a photon collides with an intense laser pulse, the dominant
QED process is non-linear Breit-Wheeler pair creation, which is
first order in the fine-structure constant $\alpha$ but all-orders
in the coupling $a_0$ to the strong background field. The probability
per unit phase of electron-positron pair creation for a photon
with energy $\omega$ and non-linearity parameter $\chi_\gamma$ is
	\begin{equation}
	\frac{\rmd P_\pm}{\rmd\phi} = \frac{W_\pm}{2\omega_0}
		= \frac{\alpha m^2 \chi_\gamma T(\chi_\gamma)}{2 \omega_0 \omega}
	\label{eq:PairCreationRate}
	\end{equation}
where (see the appendix for details) we follow~\citet{Erber} and adopt the approximation
$T(\chi_\gamma) \simeq \tfrac{0.16}{\chi_\gamma} K_{1/3}^2(\tfrac{4}{3\chi_\gamma})$,
with $K_\nu(x)$ a modified Bessel function of the second kind. The
probability~\cref{eq:PairCreationRate} is strongly suppressed for $\chi_\gamma \ll 1$.

We will determine the probability that a photon pair-creates when colliding
with an intense laser pulse, $P_\pm$, in the following way. Starting from
\cref{eq:PairCreationRate} we integrate over $\phi$ and use a saddle-point
approximation to determine the contribution to the pair creation probability
at each local maximum, calling this $P_i$. Then given $P_\pm = \sum_i P_i$ we
replace the sum over $i$ with an integral and evaluate it using another
saddle-point approximation.

Let $\phi_i$ be the phases at which the wave amplitude is (locally) maximized
and $\chi_i \equiv \chi_\gamma(\phi_i)$ the associated photon non-linearity parameter.
Then the contribution to the probability from phases near $\phi_i$ is, using
\cref{eq:PairCreationRate},
	\begin{equation}
	P_i  =
		\frac{0.16 \sqrt{3\pi} \alpha m^2}{\sqrt{8} \omega_0 \omega \sqrt{-\chi_i''}}
		\left[
			\frac{\chi_i^2 K_{1/3}^5\!\left(\frac{4}{3\chi_i}\right)}
			{K_{2/3}\!\left(\frac{4}{3\chi_i}\right) + K_{4/3}\!\left(\frac{4}{3\chi_i}\right)}
		\right]^{1/2}
	\end{equation}
where $\chi_i'' \equiv \chi''_\gamma(\phi_i)$.
In a monochromatic plane wave, or a pulse with sufficiently slowly varying
envelope, we have $\phi_i = \pi/2 + i \pi$ and
$\chi_i'' = -2 a_0 \omega_0 \omega / m^2$. Provided that $n$, the number of
pulse cycles corresponding to FWHM duration, satisfies $n>2$, we can
use these relations to determine $P_i$ as an analytical function of the index $i$.
To evaluate the sum, we replace $\sum_i \rightarrow \int \rmd i$ and perform
the integration using the Laplace method, noting that the dominant contribution
arises for $i = -1/2$, at the pulse centre.

%\begin{widetext}
%We find that the total probability for pair-creation when a photon with energy
%$\omega$ collides with a linearly-polarised laser pulse that has strength parameter
%$a_0$, frequency $\omega_0$ and (FWHM) number of cycles $n$ is
%	\begin{align}
%	P_\pm &= 1 - \exp(-\tau_\pm)
%	\\
%	\tau_\pm &\simeq
%		\alpha a_0 n \aux\!\left(\frac{2 a_0 \omega_0 \omega}{m^2}\right)
%	\label{eq:PairProbability}
%	\\
%	\aux(\chi) &=
%		\frac{ 85.1 \chi^{2/3} \Ai(x)^3 }
%		{\sqrt{2^{4/3} (7\chi^2 - 16) \Ai(x)^2 - 52\cdot2^{1/3} \chi^{4/3} \Ai(x) \Ai'(x) + 80 \chi^{2/3} \Ai'(x)^2}},
%	&
%	x \equiv (2/\chi)^{2/3}
%	\label{eq:K}
%	\end{align}
%where $\Ai$ and $\Ai'$ are respectively the Airy function and its first derivative.
%\end{widetext}

We find that the total probability for pair-creation when a photon with energy
$\omega$ collides with a linearly-polarized laser pulse that has strength parameter
$a_0$, frequency $\omega_0$ and (FWHM) number of cycles $n$ is
	\begin{equation}
	P_\pm \simeq
		\alpha a_0 n \aux
		\!\left(\frac{2 a_0 \omega_0 \omega}{m^2}\right)
	\label{eq:PairProbability}
	\end{equation}
where we have introduced an auxiliary function $\aux$ that is a function
only of the photon non-linearity parameter. $\aux$ may be expressed
analytically in terms of Airy functions, but as the underlying rate is
being treated approximately, we introduce the following functional fit for
compactness:
%	\begin{equation}
%	\aux(x) =
%		\frac{K^2_{1/3}\!\left(\frac{4}{3 x}\right)}{2(1+0.198 x^{2/3}+0.437 x)}.
%	\label{eq:K}
%	\end{equation}
%This fit is accurate to the analytical expression to within 0.5\% for
%$0.1 \leq \chi \leq 100$ and to within 2\% for all $\chi_\gamma < 1000$.
	\begin{equation}
	\aux(x) =
		\frac{0.453 K^2_{1/3}\!\left(\frac{4}{3 x}\right)}{1 + 0.145 x^{1/4} \ln(1 + 2.26 x) + 0.330 x}.
	\label{eq:K}
	\end{equation}
This fit is accurate to the analytical expression to within 1\%.

	\begin{figure}
	\subfloat[]{\label{fig:pc-a}\includegraphics[width=\linewidth]{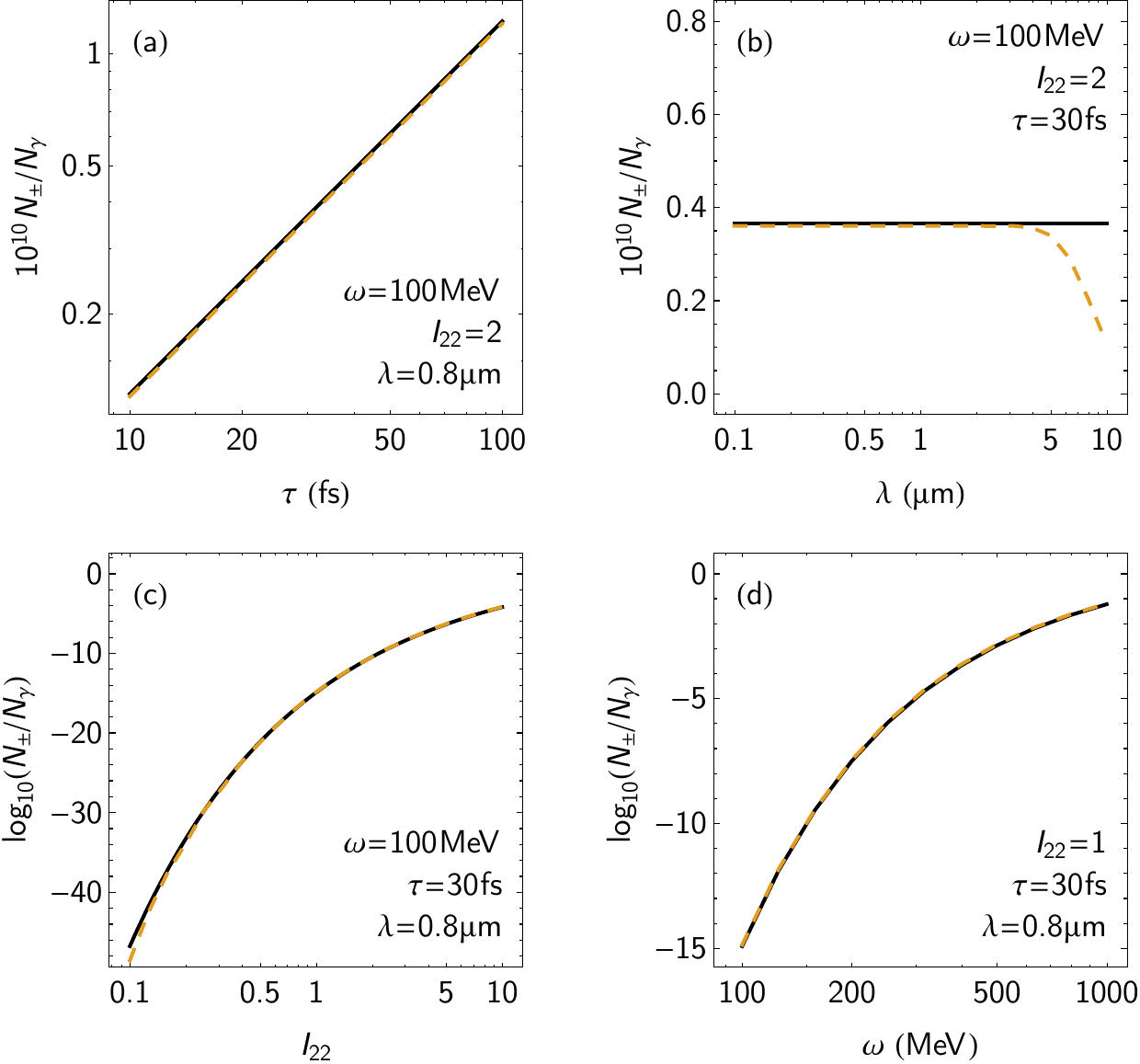}}
	\subfloat[]{\label{fig:pc-b}}
	\subfloat[]{}
	\subfloat[]{}
	\caption[Positron yields]
			{Number of positrons $N_+$ produced in a collision between a beam of
			$N_\gamma$ photons with energy $\omega$ and a linearly-polarized laser pulse
			that has peak intensity $I_{22}\times10^{22}\,\Wcm^{-2}$, wavelength $\lambda$
			and FWHM $\tau$. Results from our scaling law~\cref{eq:PairProbability}
			(black) are compared with numerical integration of the full pair creation
			rate (yellow, dashed).}
	\label{fig:PhotonCollisions}
	\end{figure}

We compare the scaling law \cref{eq:PairProbability} to the result of
numerical integration of \cref{eq:PairCreationRate} in \cref{fig:PhotonCollisions}.
Agreement is excellent across the full range of explored parameters: we capture the
super-exponential rise with increasing laser peak intensity and photon energy,
that the pair yield scales linearly with pulse length, and that it does not scale with
wavelength (provided that the wavelength is smaller than the pulse FWHM -- the drop
in positron yield in \cref{fig:pc-b} for $\lambda \gtrsim 5\,\micron$ is an effect
of the carrier phase). This is consistent with a complete calculation from
strong-field QED of the pair creation probability by \citet{Meuren}, which
concluded that $P_\pm$ scales linearly with $a_0$ at constant $\chi$ and
approximately linearly with $n$ for $a_0 \gg 1$.

The positron yield predicted by \cref{eq:PairProbability} always increases with the laser
amplitude $a_0$, the pulse length $n$ and the photon energy $\omega$. Of these
three, it is the amplitude (or peak intensity) that is the most important
as there is a dependence on $a_0$ in the prefactor and within the non-linearly
increasing function $\aux(\chi_\gamma)$.

At the high laser intensities necessary to probe QED effects, one method to
explore positron generation which guarantees overlap between the seed photons
and the laser pulse is to use the following `two stage' process. First,
a high-energy electron beam is collided with the pulse in order to generate
high-energy photons (within the pulse) by non-linear inverse Compton scattering.
The generated photons can then go on to create pairs by interaction with the laser.
In order to estimate the number of positrons produced in this configuration,
we now discuss a scaling law for the spectrum of photons produced in
non-linear Compton scattering.

\section{Gamma ray production by an electron beam}
\label{sec:grp}

\subsection{The effective centre of the laser pulse}
\label{sec:phi_c}
As the electron propagates through the intense laser pulse it loses
energy through the emission of radiation. When $\chi \gtrsim 0.1$ this emission must be
treated quantum-mechanically, as then the energy
of a single photon can be a significant fraction of the electron
energy. The following quantum corrections must be included
for our results to be predictive: the reduction in the average radiated
power by a factor $g(\chi)$~\cite{SokolovTernov}, the explicit form of which
will be given below, and the stochasticity of the emission
process~\cite{ShenWhite,Duclous,Harvey}.
The former arises because classical theory fails to preclude the
emission of photons with more energy than the electron; correcting the shape
of the spectrum to guarantee $\omega < \gamma m$ alters the scaling of
the radiated power from $\chi^2$ to $\chi^2 g(\chi)$.
%The latter means that $\gamma(\phi)$ varies
%from electron to electron, so the dynamics are determined by the evolution
%of a probability distribution $f(\gamma,\phi)$.

Here we consider typical behaviour, in the sense that the energy
loss, gamma spectra and positron yields we predict always implicitly refer to
those quantities averaged over an ensemble of electrons with the same
initial $\gamma$. Therefore the most important of the two corrections is
the factor $g(\chi)$, and we adopt a \emph{semi-classical} approach with
a modified, but deterministic, equation of motion. Furthermore we neglect
energy gain from the laser fields, requiring $\gamma \gg a_0$, such that
the evolution of the electron $\gamma(\phi)$ is determined only by radiative
losses:
	\begin{equation}
	\frac{\rmd \gamma}{\rmd \phi} = \frac{\Power}{m} =
		\frac{\alpha m \chi^2 g(\chi)}{3 \omega_0}
	\label{eq:EoM}
	\end{equation}
where $\Power$ is the power radiated per unit phase and
$g(\chi) \simeq [1+4.8(1+\chi)\ln(1+1.7\chi)+2.44\chi^2]^{-2/3}$~\cite{BKS}.

As $\chi \propto |\sin\phi|$, \cref{eq:EoM} contains an overall fluctuating factor
$\sin^2\phi$.\
%\footnote{For $\chi \gg 1$, the factor would be $\sin^{2/3}\phi$ and
%so averaging over phase would introduce a factor 0.72 rather than 0.5.}
The most important phase dependence is the envelope, so we average
over this fast oscillation, introducing an overall factor of $\tfrac{1}{2}$
to \cref{eq:EoM}. Hereafter $\chi$ will refer to the
\emph{envelope} of the electron's $\chi(\phi)$ such that
	\begin{equation}
	\chi =
		\frac{2 \gamma(\phi) a_0 \omega_0}{m}
		\exp\!\left( -\frac{\lnTwo \phi^2}{2 \pi^2 n^2} \right).
	\label{eq:ChiDefinition}
	\end{equation}
Differentiating \cref{eq:ChiDefinition} with respect to $\phi$ lets
us determine the phase $\phi_c$ at which $\chi$ is maximized.
This will prove particularly significant,
as it is where the radiated power is greatest and where the highest-energy
photons are emitted. Let $\chi_c \equiv \chi(\phi_c)$, which
satisfies the following closed relation:
	\begin{equation}
	[\chi_c^2 g(\chi_c)]^2 =
		\frac{72 \lnTwo}{\pi^2 \alpha^2}
		\left(\frac{\gamma \omega_0}{n m}\right)^2
		\ln\!\left( \frac{2 \gamma a_0 \omega_0}{m \chi_c} \right).
	\label{eq:SolveForChi}
	\end{equation}
This defines $\phi_c$ through \cref{eq:ChiDefinition}. It seems we have
made little progress though, as both $\chi_c$ and $\phi_c$ carry a dependence
on $\gamma$, the electron energy at $\phi_c$, which we do not know
\emph{a priori}. However, the presence of the correction factor $g(\chi)$
means that \cref{eq:SolveForChi} has a remarkable property: to a
good approximation, it implies that $\chi_c$ scales linearly with $\gamma$
and therefore that $\phi_c$ is independent of $\gamma$. In other words,
$\phi_c$ depends only upon the laser pulse parameters.

	\begin{figure}
	\includegraphics[width=0.8\linewidth]{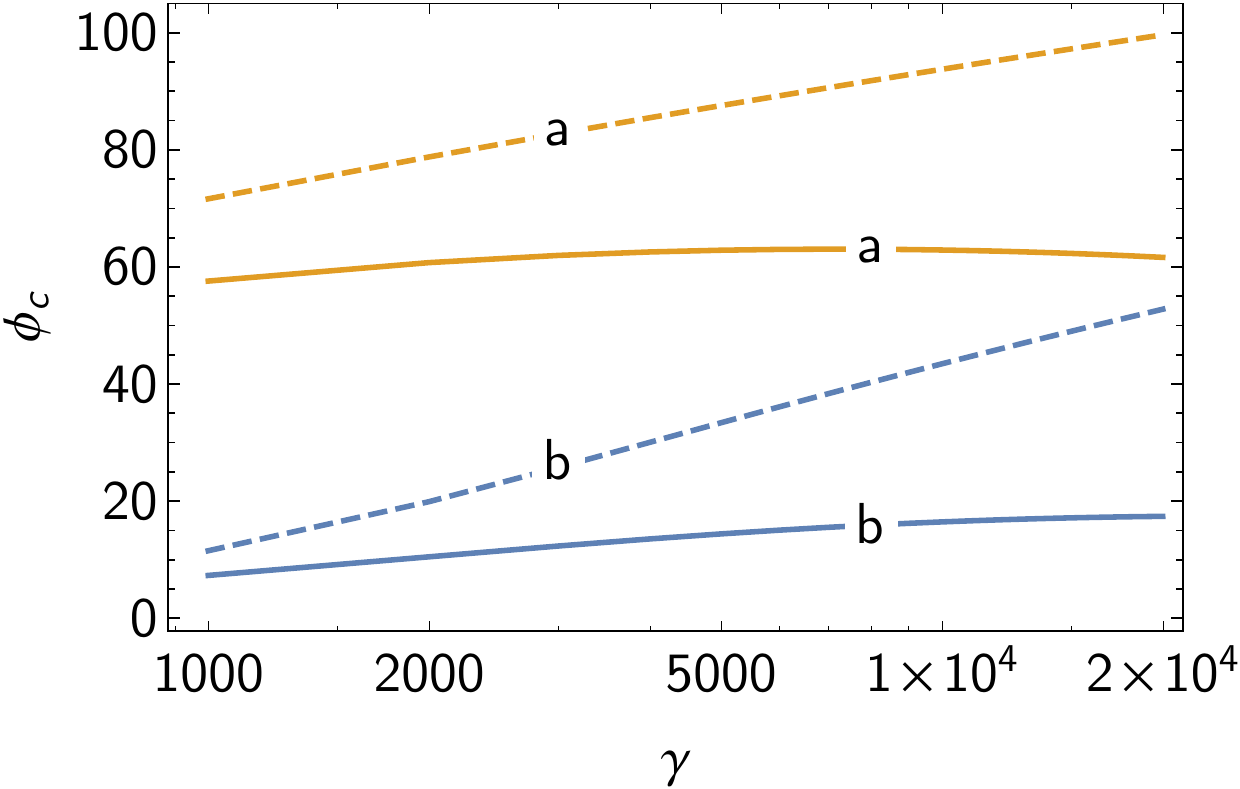}
	\caption[Critical phase]
		{$\phi_c$, the phase at which $\chi$ is maximized,
		as given by \cref{eq:ChiDefinition,eq:SolveForChi}
		for electrons colliding with laser pulses that have
		FWHM 30~fs, wavelength 800~nm (1.55~eV) and peak intensity
		(a) (yellow) $10^{23}\,\Wcm^{-2}$ and (b) (blue) $10^{21}\,\Wcm^{-2}$.
		Solid lines are calculated including $g(\chi)$; dashed
		lines have been calculated in the classical limit
		$g(\chi) = 1$.}
	\label{fig:CriticalPhase}
	\end{figure}

In \cref{fig:CriticalPhase} we show the $\phi_c$ predicted by
\cref{eq:ChiDefinition,eq:SolveForChi} for various $\gamma$ and fixed
laser pulses. It does appear that $\phi_c$ is independent of the
chosen $\gamma$ to a good approximation. To demonstrate that the origin of this effect is the
inclusion of $g(\chi)$, we also show $\phi_c$ for the classical condition
$g(\chi) = 1$. In this case, by contrast, $\phi_c$ increases with
increasing $\gamma$. Let us justify this phenomenon by differentiating
\cref{eq:SolveForChi} with respect to $\gamma$ to study the quantity
$\partial \ln \chi_c/\partial \ln \gamma$. We find that
	\begin{multline}
	\left[ \frac{2}{\chi_c} + \frac{\partial\ln g(\chi_c)}{\partial\chi_c} \right]
	\frac{\partial \chi_c}{\partial \ln \gamma} - 1
	%\gamma \frac{\partial\chi_c}{\partial\gamma}
	= \\
		\frac{A \gamma^2}{2 \chi_c^4 g(\chi_c)^2}
		%\frac{A}{2}
		%\left(\frac{\gamma}{\chi_c^2 g(\chi_c)}\right)^2
		\left( 1 -
			\frac{\partial \ln \chi_c}{\partial \ln \gamma}
			%\frac{\gamma}{\chi_c} \frac{\partial\chi_c}{\partial\gamma}
		\right)
	\label{eq:ChiGammaScaling}
	\end{multline}
where $A = 72\lnTwo \omega_0^2 / (\pi \alpha n m)^2$ is the coefficient of $\gamma^2$ on the right-hand side
of \cref{eq:SolveForChi}. The factor in square brackets on the LHS
of the above relation bears study. Classically (or, equivalently, in
the limit $\chi_c \ll 1$) it becomes $2/\chi_c$; combining this with the
fact that $A \ll 1$ we find that $\chi_c$ scales approximately
as $\gamma^{1/2}$ for $\chi_c\ll 1$.
%In the limit $\chi_c \rightarrow \infty$, it becomes
%$2/(3\chi_c)$ instead, in which case $\chi_c$ scales approximately
%as $\gamma^{3/2}$.
%However, for $\chi_c \gtrsim 0.1$, it is very close to $1/\chi_c$
%and therefore $\chi_c \propto \gamma$, $\phi_c \neq \phi_c(\gamma)$
%as suggested in \cref{fig:CriticalPhase}.
However, for $\chi_c \gtrsim 0.1$, it is very close to $1/\chi_c$
and therefore $\chi_c \propto \gamma$, giving $\mathrm{d}\phi_c/\mathrm{d}\phi \simeq 0$
as suggested in \cref{fig:CriticalPhase}.

This linear scaling only holds for `reasonable' values of $\chi_c$,
but we emphasize that because our results depend on the locally
constant field and rigid beam approximations, we begin by assuming
$\gamma \gg a_0 \gg 1$, and so for all realistic laser--electron-beam
collisions we have $\chi \gtrsim 0.1$.
We may therefore replace $\gamma$ in \cref{eq:ChiDefinition,eq:SolveForChi}
with $\gamma_0$, the Lorentz factor of the electron before the collision.
$\phi_c$ is still the phase where $\chi$ is maximized and $\chi_c$ becomes
the $\chi$ of an electron that has reached that phase without losing energy.
This is possible in the quantum radiation reaction regime due to straggling
(quenching)~\cite{ShenWhite,Harvey}.

	\begin{figure}
	\includegraphics[width=0.9\linewidth]{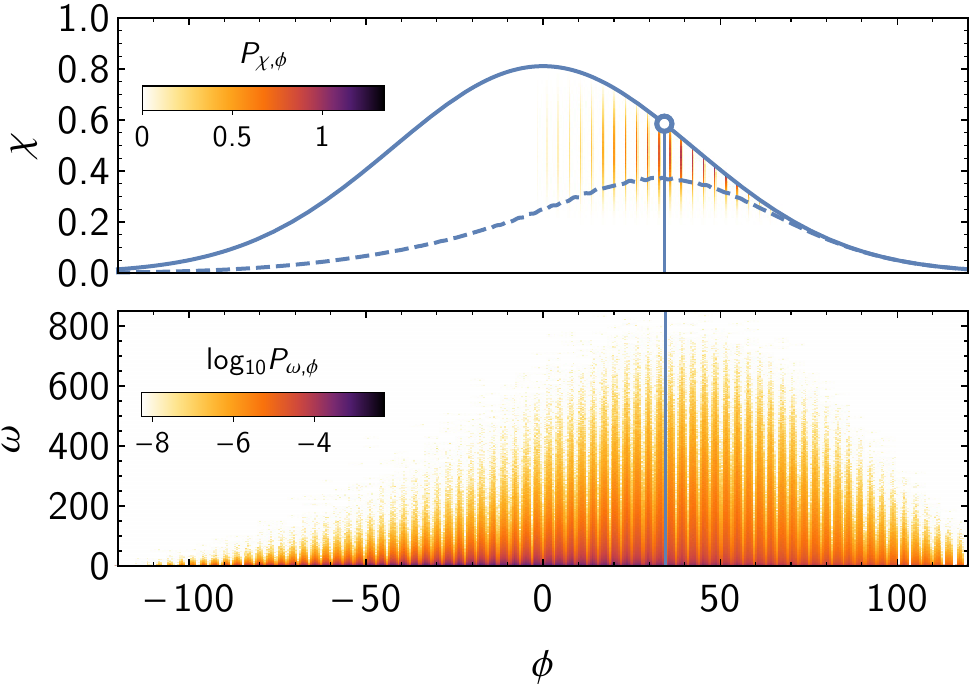}
	\caption[Maximum $\chi$]
		{Upper panel: (colour scale) the probability density $P_{\chi,\phi}$
		that a stochastically radiating electron reaches a maximum
		quantum parameter $\chi$ at phase $\phi$, (blue, solid) the
		$\chi$ of an electron that loses no energy, (blue, dashed)
		the $\chi$ of an electron that loses energy according to
		\cref{eq:EoM} and (circle) the $\chi_c$ and $\phi_c$ given by
		\cref{eq:ChiDefinition,eq:SolveForChi}.
		Observe that the region of maximum emission probability is
		correctly identified by the predicted $\phi_c$.
		Lower panel: (colour scale) the probability density $P_{\omega,\phi}$
		that a 	photon is emitted with energy $\omega$ at phase $\phi$ and
		(vertical line) $\phi_c$. (See text for collision parameters.)}
	\label{fig:ChiMax}
	\end{figure}

\Cref{fig:ChiMax} compares the predicted $\phi_c$ and $\chi_c$ to the
results of a single-particle Monte Carlo simulation of quantum radiation
reaction. The initial energy of the electron is 1\,GeV and the laser pulse
has wavelength $0.8\,\micron$, FWHM $30\,\fs$ and peak intensity
$10^{22}\,\Wcm^{-2}$ ($\gamma_0 = 1957$, $\omega_0 = 1.55\,\eV$, $n = 11.2$
and $a_0 = 68.3$). For each electron we track the maximum $\chi$
experienced along its trajectory, as well as the phase at which this
occurred. The probability density $P_{\chi,\phi}$ that an electron reaches
$\chi$ at phase $\phi$ is plotted in the upper panel of \cref{fig:ChiMax};
to aid the eye, the $\chi(\phi)$ of a non-radiating and semi-classically
radiating electron are plotted as well. We see that $\phi_c$ accurately
captures the point at which the electron $\chi$ is maximized, both in
the semi-classical and stochastic cases. Comparison with the probability
density $P_{\omega,\phi}$ that a photon is emitted with energy $\omega$
at phase $\phi$, plotted in the lower panel, shows that $\phi_c$ also
characterizes the region where the highest energy photons are emitted.

\subsection{Energy loss of the electron beam}
\label{sec:rr}

We now derive a scaling law for the energy loss of an electron in the
quantum-radiation-reaction--dominated regime. Of course, we could
simply solve \cref{eq:EoM} given the pulse parameters, but as
$g(\chi)$ has no simple analytical form, those results would
necessarily be numerical. Instead we use the results of \cref{sec:phi_c}
to guide us toward an approximate, but analytical, scaling law.
We expect that $\chi$ and the radiated power are strongly peaked in the region
around $\phi = \phi_c$, so we will use the Laplace method (i.e. the saddle
point approximation) to estimate the radiated energy in the absence
of recoil. Then we will employ the single photon recoil correction
$\Omega \rightarrow \Omega / (\gamma m - \Omega)$ to obtain a
recoil-corrected estimate of both $\chi_c$ and the final electron
energy.

The Laplace method for the integral $\int\!\Power(\phi)\,\rmd\phi$
effectively replaces the integrand with a Gaussian with peak
$\Power_c$ and variance $\sigma^2 = -\Power_c/\Power''_c$, these being
evaluated at the point $\phi_c$ where $\Power'$ vanishes. Here
primes denote differentiation with respect to $\phi$.
Then the integral is $[2\pi \Power_c^3/(-\Power''_c)]^{1/2}$. We have
that
	\begin{equation}
	\Power_c = \frac{\alpha m^2 \chi_c^2 g(\chi_c)}{6 \omega_0}
	\end{equation}
using the results of \cref{sec:phi_c}. The second derivative
	\begin{equation}
	\Power''_c =
		\Power_c
		\left[ \frac{2}{\chi_c} + \frac{\partial\ln g(\chi_c)}{\partial\chi_c} \right]
		\left. \chi'' \right|_{\phi=\phi_c}
		%\left. \frac{\partial^2 \chi}{\partial \phi^2}\right|_{\phi=\phi_c}
	\label{eq:SecondDvPower}
	\end{equation}
where
	\begin{equation}
	\left. \chi'' \right|_{\phi=\phi_c} =
		-\frac{\lnTwo \chi_c}{\pi^2 n^2}
		\left[
			1 +
			2 \ln\!\left( \frac{2\gamma_0 a_0 \omega_0}{m \chi_c} \right)
		\right].
	\label{eq:SecondDvChi}
	\end{equation}
\Cref{eq:SecondDvPower} contains the same factor in square brackets
as \cref{eq:ChiGammaScaling}; as before, we replace it with $1/\chi_c$.
Then we find that the radiated energy (in the absence of recoil) is
%	\begin{equation}
%	\Omega =
%		\frac{\pi^{3/2} \alpha m^2 n}{3 \sqrt{2\lnTwo} \omega_0}
%		\frac{\chi_c^2 g(\chi_c)}{\sqrt{1 + 2 \ln\!\left(\frac{2\gamma_0 a_0 \omega_0}{m \chi_c}\right)}}.
%	\end{equation}
	\begin{equation}
	\Omega =
		\sqrt{2\pi} \gamma_0 m
		\left[
			\frac{2\ln\!\left( \frac{2\gamma_0 a_0 \omega_0}{m \chi_c} \right)}
					{1+2\ln\!\left( \frac{2\gamma_0 a_0 \omega_0}{m \chi_c} \right)}
		\right]^{1/2}.
	\label{eq:RadEnergyNoRecoil}
	\end{equation}
The argument of the logarithms is always $\geq 1$; the equality would
correspond to $\phi_c = 0$ and $\chi_c$ taking on its largest possible value
at the pulse centre. Therefore $0 \leq \Omega \leq \sqrt{2\pi} \gamma_0 m$.

Let $\Omega_\text{rr}$ be the total energy emitted in photons when we do account for
the electron recoil, i.e. radiation reaction. Were we to assume that only
one photon is emitted, the first order correction would give
$\Omega_\text{rr} = \Omega / (1 + \Omega / \gamma_0 m)$~\cite{Schwinger,Lindhard}.
However, as the electron emits many photons, this is not a very good approximation. We will
be guided instead by the fact that the radiated energy should be approximately
symmetric around the point $\phi = \phi_c$. This is exactly true for the
Laplace method because the fitted Gaussian is centred at $\phi = \phi_c$.
It will still be true after we account for recoil because $\phi_c$ is
independent of $\gamma$ and must still mark the point of maximum radiated
power. We have, by our argument for the symmetry of the radiated energy,
that the electron loses $\Omega/2$ during the interval
$\phi > \phi_c$; therefore its Lorentz factor and recoil-corrected
$\chi$ at $\phi_c$ satisfy
	\begin{equation}
	\frac{\gamma_c}{\gamma_0} =
		\frac{\chicrr}{\chi_c} \simeq
		\left(1 + \frac{\Omega}{2 \gamma_0 m}\right)^{-1}.
	\label{eq:MidPointGamma}
	\end{equation}
Repeating the process for the interval $\phi < \phi_c$, the electron final
Lorentz factor is then
	\begin{equation}
%	\gamma_\mathrm{f} =
%		\frac{8 \gamma_0^4 m^3}{(2 \gamma_0 m + \Omega)(4 \gamma_0^2 m^2 + 2 \gamma_0 m + \Omega^2)}.
	\gamma_\mathrm{f} \simeq
		\frac{2\gamma_0 m - \Omega}{2\gamma_0 m + \Omega}
		\gamma_0.
	\end{equation}
This is positive only if $\Omega < 2 \gamma_0 m$; as we saw earlier,
$\Omega$ is bounded by $\sqrt{2\pi}\gamma_0m \simeq 2.51\gamma_0 m$.
Nevertheless, as we seek a scaling law for the photon spectrum, it is
more important that both $\gamma_c$ and $\chicrr$ are
correctly bounded by zero from below, which they indeed are.

We can also estimate the electron Lorentz factor and $\chi$ as a function
of phase, using that the radiated power and $\chi$ as functions of phase are
approximately Gaussian in form:
	\begin{align}
	\gamma(\phi) &\simeq
		\gamma_\mathrm{f} +
		\frac{\gamma_0 \Omega}{2 \gamma_0 m + \Omega}
		\left[
			1 + \erf\!\left(
				\frac{\phi-\phi_c}{\sqrt{2}\sigma}
				\right)
		\right]
	\label{eq:GammaPhase}
	\\
	\chi(\phi) &\simeq
		\frac{\chi_c}{1 + \frac{\Omega}{2 \gamma_0 m}}
		\exp\!\left( -\frac{(\phi-\phi_c)^2}{2 \sigma^2} \right)
	\label{eq:ChiPhase}
	\end{align}
where
	\begin{equation}
	\sigma^2 =
		\frac{\pi^2 n^2}{\lnTwo}
		\left[
			1 + 2 \ln\!\left( \frac{2\gamma_0 a_0 \omega_0}{m \chi_c} \right)
		\right]^{-1}.
	\end{equation}
Comparison between \cref{eq:ChiPhase} and numerical solutions to
the equation of motion \cref{eq:EoM} are given in \cref{fig:PhotonSpectra},
for experimental parameters corresponding to Gemini (\cref{fig:spec-Gemini}),
BELLA (\cref{fig:spec-Bella}),
SLAC (\cref{fig:spec-SLAC}) and ELI (\cref{fig:spec-ELI}). There is
excellent agreement with respect to the maximum $\chi$, the phase at which it
is reached, and the distance over which it is sustained.
This demonstrates the importance of accounting for radiation reaction,
as without doing so we would overestimate $\chi$.

\subsection{Photon energy spectrum}
\label{sec:PhotonSpectrum}

\Cref{fig:ChiMax} confirms that the region near $\phi_c$, where $\chi$ is maximized,
is the origin of the highest-energy photons. We propose that the spectrum may be
approximated by assuming that the electron has $\chi$ as given by \cref{eq:ChiPhase}
and energy $\gamma \simeq \gamma_0$ at this point; the latter is our way of accounting
for straggling (or quenching). As we are interested in the high-energy tail of the spectrum,
we expand the double differential rate of emission for $\chi_\gamma \sim \chi$:
	\begin{equation}
	\frac{\partial^2 N_\gamma}{\partial t \partial \chi_\gamma} \simeq
		\frac{\alpha m}{2 \sqrt{\pi} \gamma}
		\frac{\exp\!\left(-\frac{2 \chi_\gamma}{3\chi (\chi - \chi_\gamma)}\right)}
		{\sqrt{\chi - \chi_\gamma}},
	\end{equation}
which may be converted into an integral over phase and photon energy using that
$\phi = -2\omega_0t$ and $\omega/(\gamma m) = \chi_\gamma / \chi$. Then we use
the Laplace method again with the relations \cref{eq:SecondDvChi,eq:ChiPhase},
with the result that
	\begin{multline}
	\frac{\rmd N_\gamma}{\rmd \omega} \simeq
		\frac{\sqrt{3}\pi\alpha F_\text{he}}{\sqrt{2\lnTwo}}
		\frac{a_0 n}{\sqrt{E_0}}
	\\ \times
		\frac{\chicrr / \chi_0}
			{\sqrt{1 + 2\ln\!\left( \chi_0 / \chi_c \right)}}
	\\ \times
		\frac{\exp\!\left(
						-\frac{2 \omega}{3\chicrr(E_0 - \omega)}
					\right)}
			{\sqrt{3\chicrr (E_0 - \omega) + 4 \omega}}
	\label{eq:PhotonSpectrum}	
	\end{multline}
%	\begin{widetext}
%	\begin{equation}
%	\frac{\rmd N_\gamma}{\rmd \omega} \simeq
%		\frac{\sqrt{3}\pi\alpha F_\text{he}}{\sqrt{2\lnTwo}}
%		%\frac{m^2 n}{E_0^{3/2} \omega_0}
%		\frac{a_0 n}{\sqrt{E_0}}
%		%\frac{\chicrr}
%		\frac{\chicrr / \chi_0}
%			{\sqrt{1 + 2\ln\!\left(
%									%\frac{2 a_0 E_0 \omega_0}{m^2 \chi_c}
%									\chi_0 / \chi_c
%								\right)}}
%		\frac{\exp\!\left(
%						-\frac{2 \omega}{3\chicrr(E_0 - \omega)}
%					\right)}
%			{\sqrt{3\chicrr (E_0 - \omega) + 4 \omega}}
%	\label{eq:PhotonSpectrum}
%	\end{equation}
%	\end{widetext}
for $E_0 = \gamma_0 m$, $\chi_0 = 2 \gamma_0 a_0 \omega_0 / m$ and $\chicrr$
related to $\chi_c$ via \cref{eq:MidPointGamma}. Now, as not every emission
qualifies as `high-energy', this will overestimate the number of hard photons.
To account for this we multiply \cref{eq:PhotonSpectrum} by a correction
factor $F_\text{he}$ which is the ratio of the number of photons emitted
for $\phi > \phi_c$ to the total number of photons
	\begin{equation}
	F_\text{he} =
		\frac{1}{2}
		\left[
		1 - \erf\!\left( \frac{\sqrt{2\lnTwo}\phi_c}{2 \pi n} \right)
		\right].
	\label{eq:CorrectionFactor}
	\end{equation}
This works because, as we showed in \cref{sec:rr}, the electron
loses most of its energy for $\phi \simeq \phi_c$; only for larger phases can
it be said still to be `high-energy'.

	\begin{figure}
	\subfloat[]{\label{fig:spec-Gemini}\includegraphics[width=\linewidth]{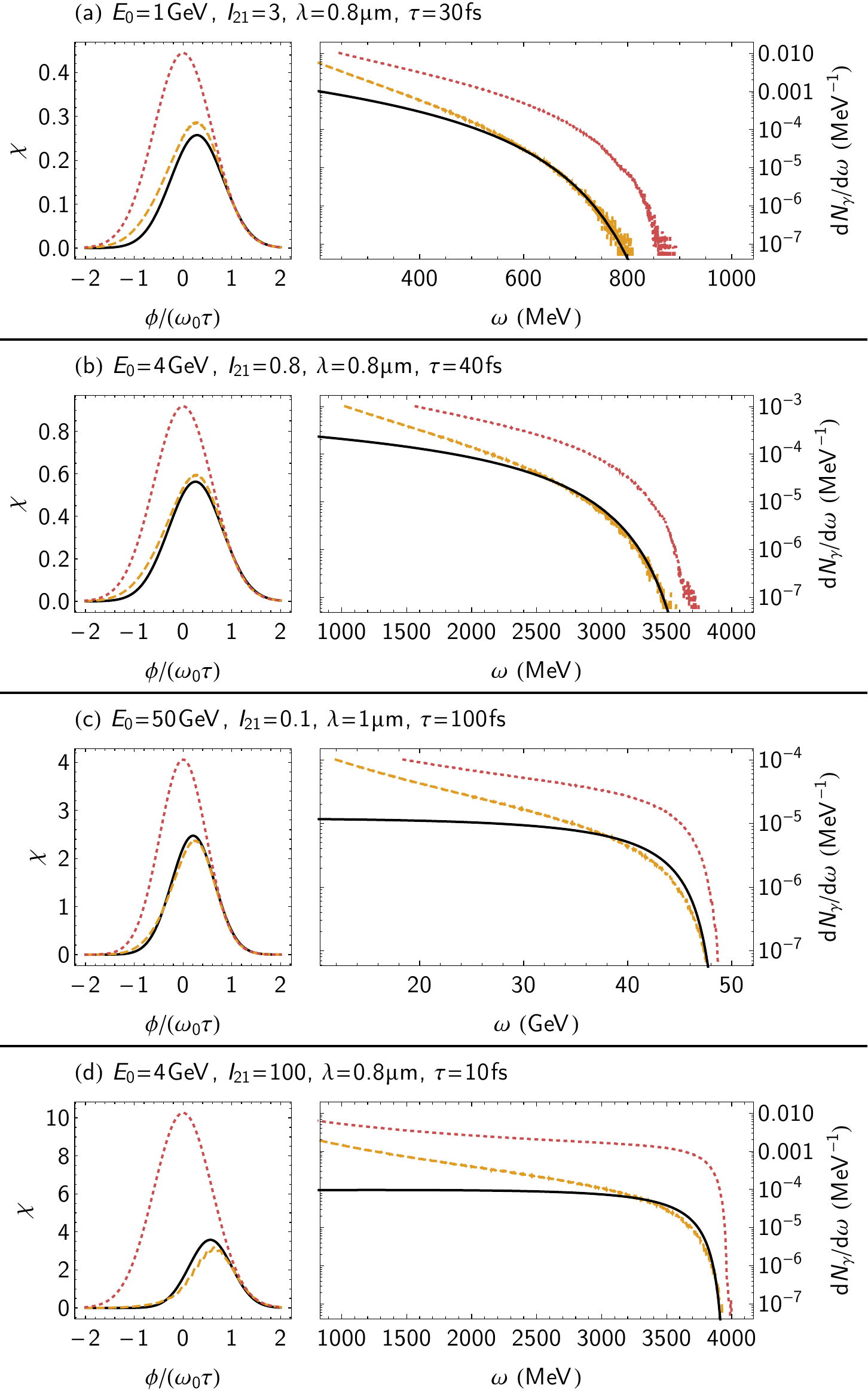}}
	\subfloat[]{\label{fig:spec-Bella}}
	\subfloat[]{\label{fig:spec-SLAC}}
	\subfloat[]{\label{fig:spec-ELI}}
	\caption[Photon spectra]
			{For a collision between an electron beam with
			energy $E_0$ and a linearly-polarized laser pulse with peak intensity
			$I_{21}\times10^{21}\,\Wcm^{-2}$, wavelength $\lambda$
			and FWHM $\tau$:
			(left panels) the electron quantum non-linearity
			parameter $\chi$ as a function of phase $\phi$, as predicted by
			\cref{eq:ChiPhase}, (yellow, dashed) solution to equation of motion
			\cref{eq:EoM}, and (red, dotted) in the absence of radiation reaction;
			and (right panels) energy spectra (normalized per electron) of the
			emitted photons, as predicted
			by (black) our scaling \cref{eq:PhotonSpectrum} and Monte Carlo simulation
			with (yellow, dashed) stochastic radiation reaction and (red, dotted)
			no radiation reaction.}
	\label{fig:PhotonSpectra}
	\end{figure}
	
We compare the predicted scalings with simulation data in \cref{fig:PhotonSpectra}
(for consistency, pair creation and therefore secondary photon emission were
disabled).
The log scaling of the vertical axes admittedly flatters the comparison, but
we find good agreement between the stochastic data and our simple scaling.
It captures both the shape of the high-energy tail, the absolute number of
photons and the reduction in both caused by radiation reaction.

\section{Positron production by an electron beam}
\label{sec:PairCreation}

\subsection{Number of pairs}
\label{ssec:ElectronBeam}

The number of positrons produced by a laser--electron-beam collision may
be determined by convolving the pair creation probability \cref{eq:PairProbability}
with the photon spectrum \cref{eq:PhotonSpectrum}. Strictly, this requires that
the contribution to the probability $P_\pm$ from the region $\phi > \phi_c$ is
negligible, as the daughter photon beam is actually created within
the laser pulse near $\phi = \phi_c$.

While the pair creation probability is always (non-linearly) increasing with increasing
photon energy, the photon number is always decreasing because of the
exponential factor in \cref{eq:PhotonSpectrum}. Therefore the probability
spectrum $P_\pm \tfrac{\rmd N_\gamma}{\rmd \omega}$ is peaked
for some $\omega_c < \gamma_0 m$. (If consider the ensemble of positrons emerging from
the laser-beam collision and look at the energy distribution of their parent
photons, $\omega_c$ will be most probable.) Let us consider the threshold regime
for pair creation, which is currently accessible, where the yield is dominated
by the conversion of the highest energy photons. Then we may expect the combination
of our scaling laws for the photon spectrum and pair probability to be predictive.

First we derive a relation for $\omega_c$. Let $S(\omega)$ be the last factor in
\cref{eq:PhotonSpectrum}, the only part that depends on the photon energy.
Then the product $P_\pm \tfrac{\rmd N_\gamma}{\rmd \omega}$ is maximized for
$\omega = \omega_c$ satisfying
	\begin{equation}
	%\frac{2 a_0 \omega_0}{m^2}
	\frac{\aux'(\chi_\gamma)}{\aux(\chi_\gamma)}
	\frac{\partial \chi_\gamma}{\partial \omega}
	 = \frac{\abs{S'(\omega_c)}}{S(\omega_c)}
	\end{equation}
where $\chi_\gamma = 2 a_0 \omega_0 \omega_c / m^2$ and $\aux$ was introduced in
\cref{eq:K}. We expect $\omega_c$ to be near $E_0$, so we take only the leading
order term in $(\gamma_0 m - \omega)$ from the RHS. The LHS depends on the
properties of $\aux$ but we find that for
$\chi_\gamma \leq 10$, $\aux'/\aux \simeq 2.75/\chi_\gamma^2$.
Therefore we have
	\begin{equation}
	\omega_c \simeq
		\gamma_0 m
		\frac{\sqrt{\frac{2 \chicrr m}{a_0 \gamma_0 \omega_0}}}
		{1 + \sqrt{\frac{2 \chicrr m}{a_0 \gamma_0 \omega_0}}}.
	\label{eq:OmegaC}
	\end{equation}

We use this point as the origin of a saddle-point approximation to the integral
$\int P_\pm \tfrac{\rmd N_\gamma}{\rmd \omega} \rmd\omega$ which will give us
the positron yield arising from a high-energy electron beam.
%We neglect the exponential correction ($P_\pm \simeq \tau_\pm$)
We take only the leading order term in $(\gamma_0 m - \omega_c)$ as before.
Leaving out the details, we find that the number of positrons produced per electron is
	\begin{multline}
	N_+ \simeq
		\frac{3 \sqrt{\pi} P_\pm (\omega_c) \chicrr }{\sqrt{2}}
		\frac{(\gamma_0 m - \omega_c)^2}{\gamma_0 m}
		\left. \frac{\rmd N_\gamma}{\rmd \omega} \right|_{\omega = \omega_c}
	\label{eq:PositronYieldE}
	\end{multline}
using the recoil-corrected $\chicrr$ from \cref{eq:MidPointGamma},
$P_\pm$ from \cref{eq:PairProbability} and the photon spectrum
from \cref{eq:PhotonSpectrum}.

	\begin{figure}
	\subfloat[]{\label{fig:py-a}\includegraphics[width=\linewidth]{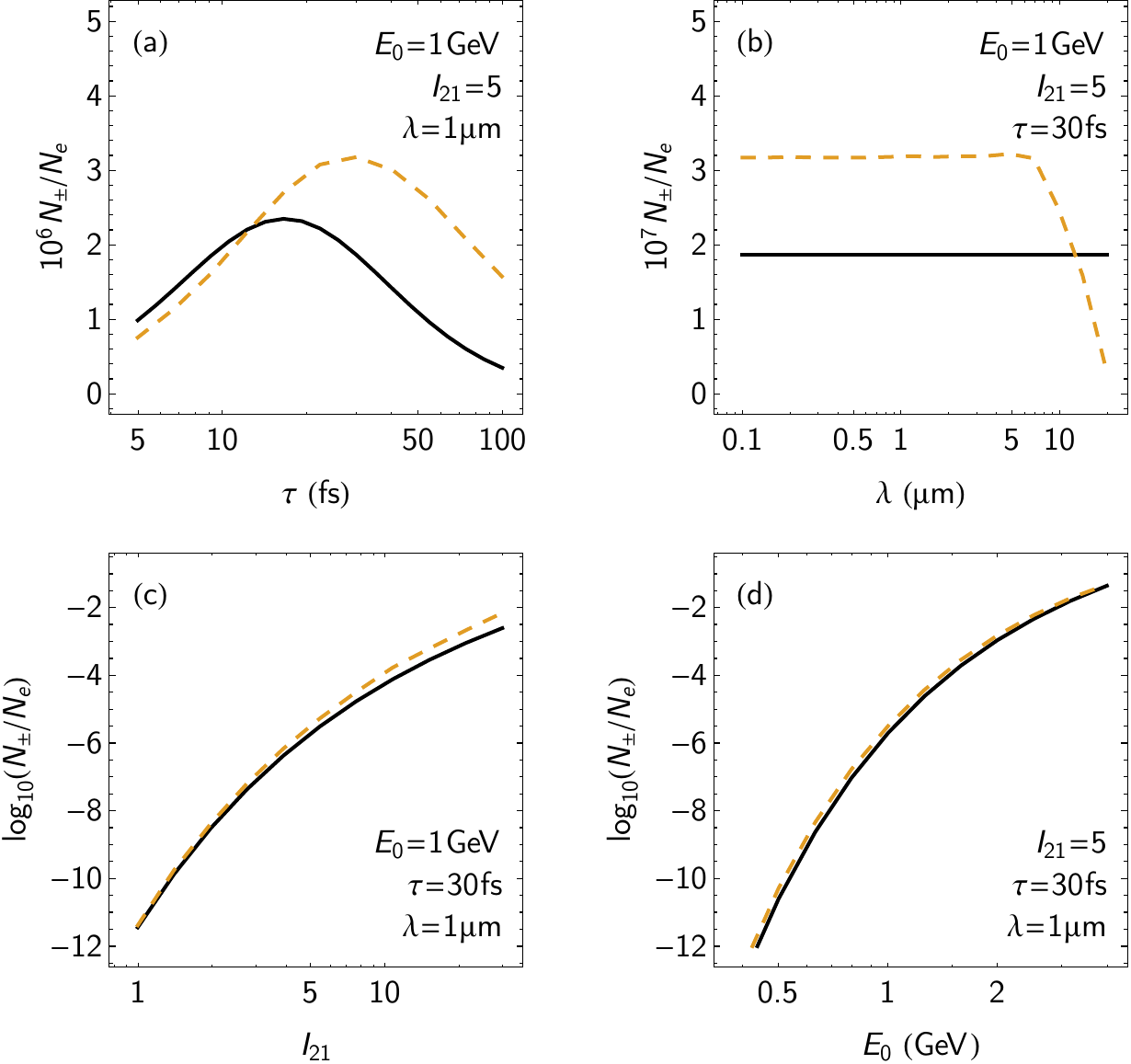}}
	\subfloat[]{}
	\caption[Positron yield]
			{Number of positrons $N_+$ produced in a collision between a beam of
			$N_e$ electrons with energy $E_0$ and a linearly-polarized laser pulse
			that has peak intensity $I_{21}\times10^{21}\,\Wcm^{-2}$,
			wavelength $\lambda$ and FWHM $\tau$.
			Results from (black) our scaling law \cref{eq:PositronYieldE}
			and (yellow, dashed) simulations using 
			the full pair creation rate.}
	\label{fig:PositronYield}
	\end{figure}
	
The results of this calculation are compared with
the yield obtained from Monte Carlo simulation in \cref{fig:PositronYield},
for collision parameters that are within the scope of present-day laser
facilities. We find that it is accurate to within a factor of 2 to 3 across
the range of explored parameters, with a tendency to underestimate the yield.
This is because the approximate spectrum \cref{eq:PhotonSpectrum}, while
accurate for the high-energy tail, underestimates the number of low to
mid-energy photons. At lower intensities, positron production is dominated
by the high-energy tail of the spectrum so our prediction is accurate. As $\chi_\gamma$
exceeds 1, pair creation is possible for photons across a wider energy
range, and our prediction will undershoot.

Increasing $\gamma$ and $a_0$ for fixed pulse length $n$
always increases the positron yield. However, for fixed $\gamma_0$ and $a_0$, we see from
\cref{fig:py-a} that there is a laser pulse length where the positron yield
is maximized. This may be understood by considering the competing factors
of $P_\pm$ and $\chicrr$ in \cref{eq:PositronYieldE}. The former
favours increasing pulse length as photon decay becomes more probable. The latter
accounts for the sensitivity of the yield to the pulse rise time (which increases
with $n$), as increased energy loss of the electron beam in the rising edge
suppresses growth of $\chi$ and consequently hard photon emission. This is
why our scaling underestimates the yield for $\tau \gtrsim 40\,\fs$: the
photon spectrum in this region is dominated by low- to mid-energy photons
which the scaling \cref{eq:PhotonSpectrum} does not capture.

\subsection{Mean positron energy}
\label{ssec:MeanPositronEnergy}

Now we consider the effect of secondary photon emission on the positron
energy. To produce large numbers of pairs we need $\chi_\gamma \gtrsim 1$,
implying a positron initial $\chi_+ \gtrsim 0.5$ which lies within the
quantum radiation reaction regime. We will assume that the positron (equivalently,
the electron) is created at $\phi = 0$ with half the energy of the parent photon.
We take this to be the most probable parent photon energy $\omega_c$, as
given by \cref{eq:OmegaC}. The procedure to determine the final energy of
the positron is similar to that outlined in \cref{sec:rr}. Both $\chi_+$ and
the radiated power are maximized at $\phi = 0$ and decrease with decreasing $\phi$.
Given this, we integrate the radiated power from $\phi = -\infty$ to
$\phi = 0$ using the Laplace method to determine radiated energy without recoil,
then apply the single-photon correction. We find that the post-collision Lorentz
factor of the positron
	\begin{equation}
	\gamma_{+} \simeq
		\frac{\omega_c}{2m}
		\left[
		1 +
			\frac{\pi^{3/2}\alpha}{3 \sqrt{2\lnTwo}}
			\frac{n a_0^2 \omega_0 \omega_c}{m^2}
			g\!\left(\frac{a_0 \omega_0 \omega_c}{m^2}\right)
		\right]^{-1}.
	\label{eq:PositronEnergy}
	\end{equation}

	\begin{figure}
	\subfloat[]{\label{fig:pe-a}\includegraphics[width=\linewidth]{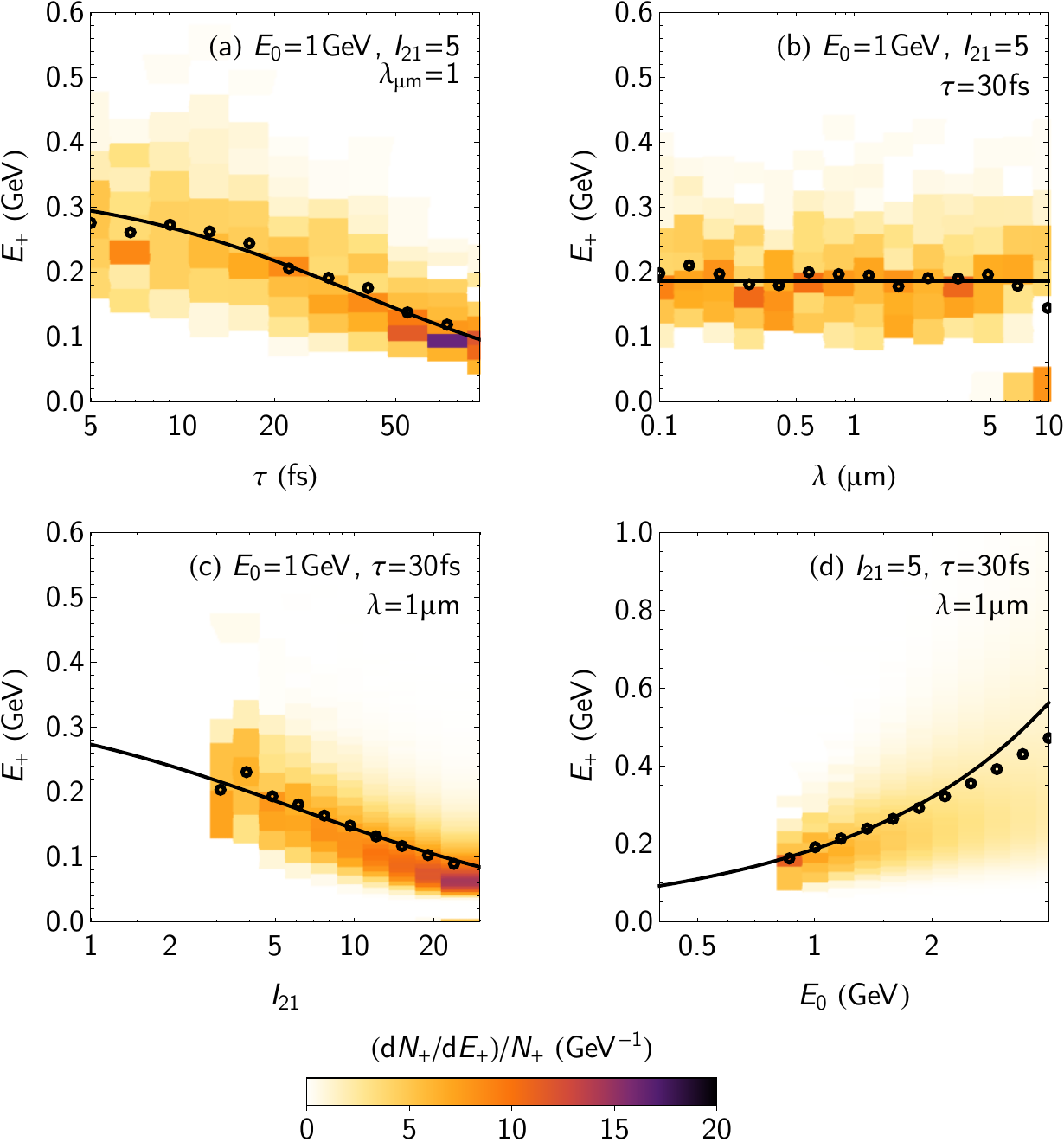}}
	\subfloat[]{}
	\subfloat[]{}
	\subfloat[]{}
	\caption[Positron spectra]
			{(colour scale) Energy spectra of positrons emerging from a collision
			between an electron
			beam with energy $E_0$ and a laser pulse with peak intensity
			$I_{21}\times10^{21}\,\Wcm^{-2}$, wavelength $\lambda$ and FWHM $\tau$,
			(black lines) the characteristic energy predicted by
			\cref{eq:OmegaC,eq:PositronEnergy} and (black circles) the mean
			energy of the simulated spectra.}
	\label{fig:PositronSpectra}
	\end{figure}

Comparison with Monte Carlo simulation, shown in \cref{fig:PositronSpectra},
shows that this scaling law accurately predicts the mean positron final energy.
Nevertheless, we see it breaking down for $\gamma_0 m$ larger than a GeV, as the initial positron
spectrum for $\chi_\gamma > 1$, while still symmetric around $m\gamma_+ =
\omega_c/2$, is much broader and stochastic effects are more pronounced.
The positron spectrum is unchanged as the wavelength increases until 
$n = c \tau / \lambda \gtrsim 2$, at which point carrier phase effects
become significant. The laser pulse we consider has phase dependence
$\sin\phi$, so this means the pair creation is switched off as the pulse
FWHM shrinks. In general the width of positron spectrum increases with
increasing electron beam energy, and decreases with increasing pulse
intensity and duration.

\subsection{Prospects for experimental observation}
\label{ssec:Expt}

	\begin{figure*}
	\subfloat[]{\label{fig:pt-a}\includegraphics[width=0.4\linewidth]{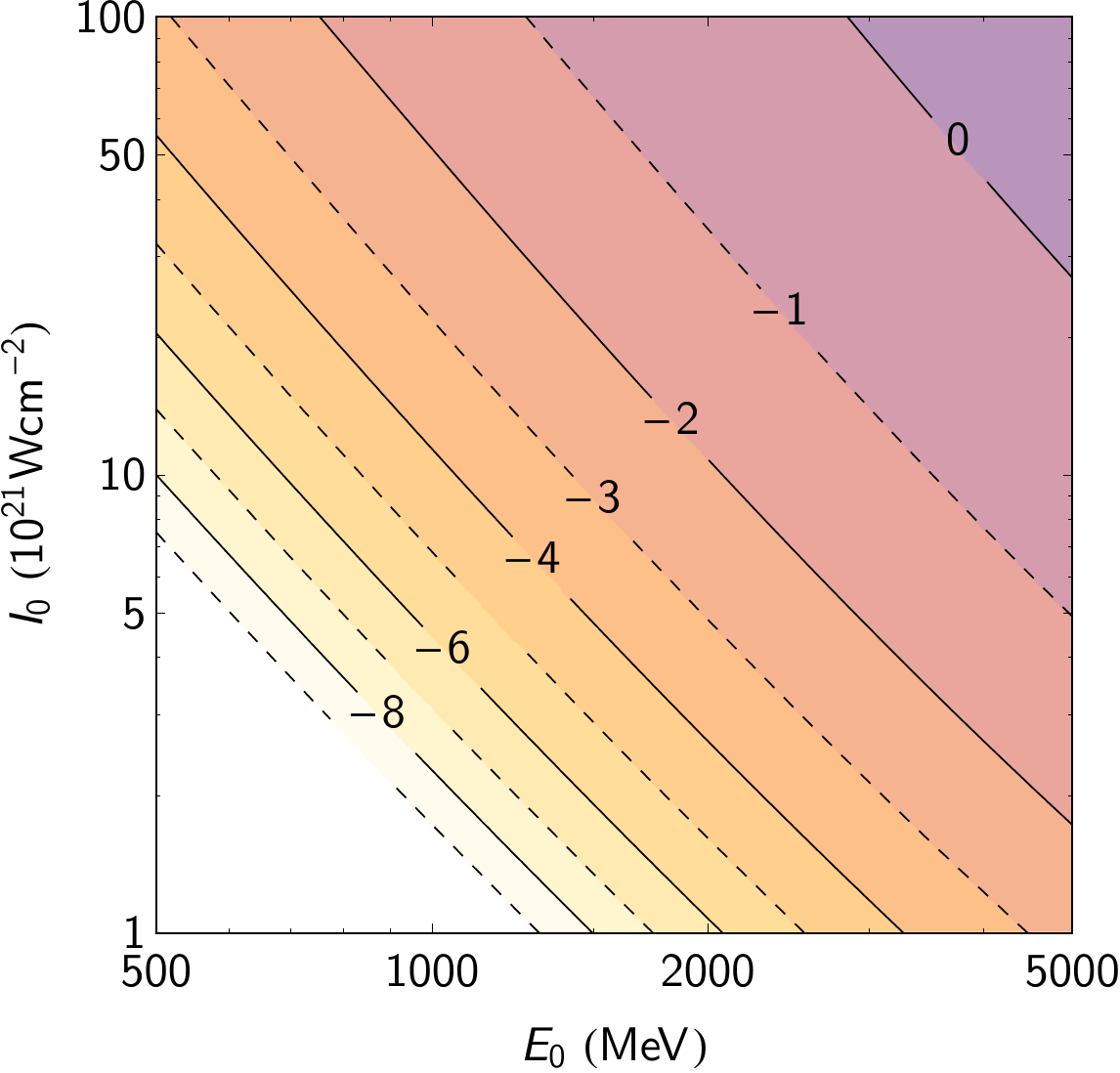}}
	%\vspace{0pt}
	\subfloat[]{\label{fig:pt-b}\includegraphics[width=0.4\linewidth]{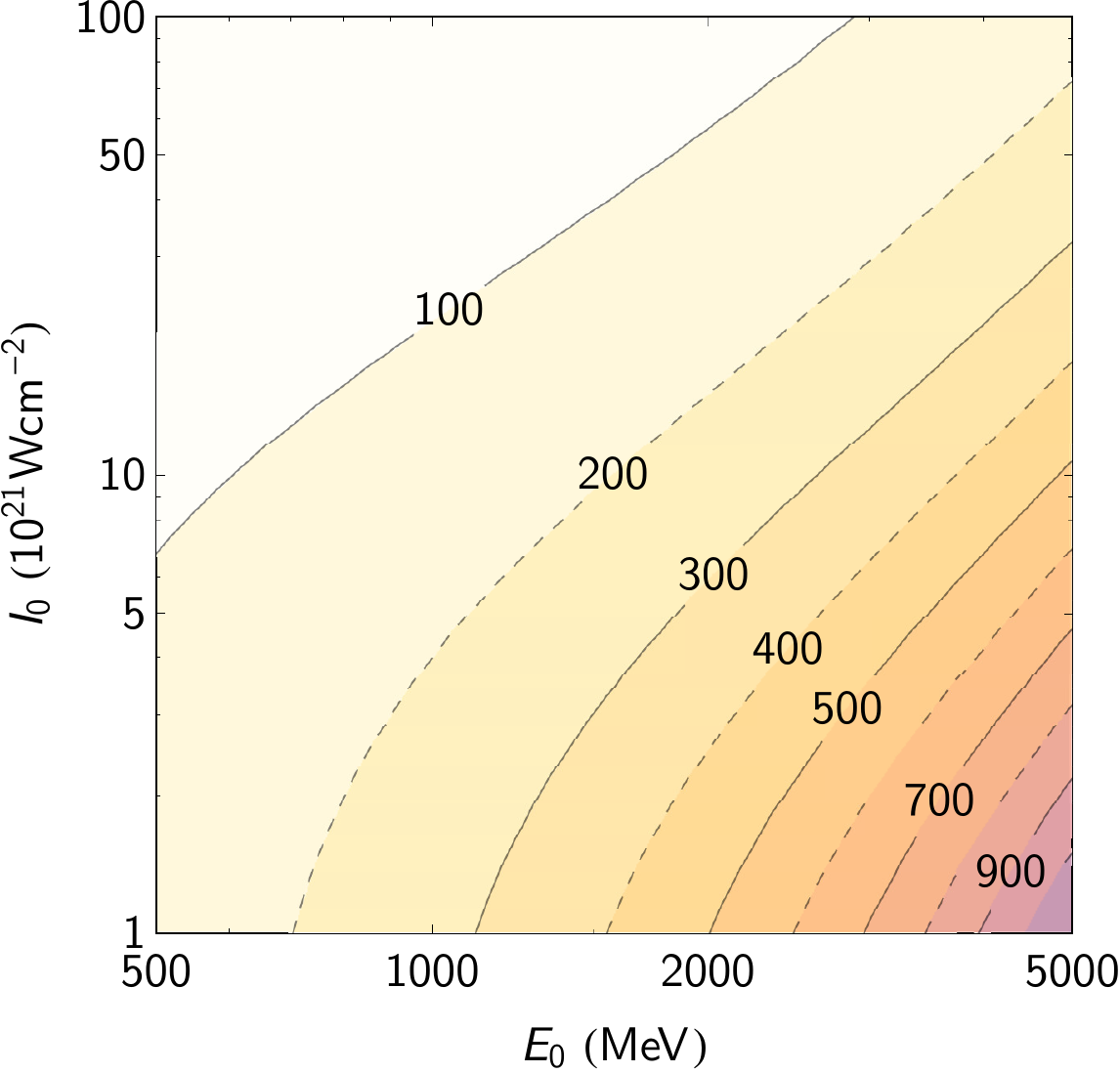}}
	\caption[Experimental parameters]
			{(left) $\log_{10}$-scaled yield (per electron)
			and (right) typical energy in MeV of positrons produced
			in the collision of an electron beam with energy $E_0$
			and a laser pulse with peak intensity $I_0$, wavelength $1\,\micron$
			and FWHM $30\,\fs$, as predicted by \cref{eq:PositronYieldE}.}
	\label{fig:PositronTotal}
	\end{figure*}

We have seen that the optimal pulse length for pair creation from
GeV electron beams and laser pulses with intensity in the high $10^{21}\,\Wcm^{-2}$
is approximately $30\,\fs$, which is close to the characteristic duration
of current high-intensity lasers. Therefore we show in \cref{fig:PositronTotal}
the number and characteristic energy of positrons produced in such a collision,
taking the wavelength of the light to be $1\,\micron$ and the FWHM of the pulse
to be $30\,\fs$.
The positron yield increases substantially with both increasing electron beam
energy and laser intensity. Laser wakefields typically accelerate bunches
of charge $100\,$pC, implying that to produce more than
100 positrons in a single shot requires a laser intensity $I_0$ and beam
energy $E_0$ that satisfy
	\begin{equation}
	\left(\frac{I_0}{10^{21}\,\Wcm^{-2}}\right)
	\left(\frac{E_0}{2\,\text{GeV}}\right)^2 \gtrsim 1.
	\label{eq:Condition}
	\end{equation}
This condition can be met in present-day high-intensity laser facilities,
where we expect measurable quantities of $>100\,\MeV$ positrons to be produced.
However, to be confident that this is the case, we extend our results
to account for the fact that both laser and electron beam
have finite size.

Consider a beam of $N_{e,\mathrm{b}}$ electrons with spherically symmetric Gaussian
charge density (size $R$), offset from the laser axis by a distance $\Delta$.
Without loss of generality we may choose that offset to be in the $x$-direction.
The peak laser intensity each electron encounters depends upon
that electron's spatial and temporal offset from the laser focus. Let $(x,y)$ 
be the position at which an individual electron encounters the peak of the
laser pulse. Then the effective $a_0$ of the pulse for that electron becomes
	\begin{equation}
	a \simeq a_0
		\exp\!\left( -\frac{x^2 + y^2}{w_0^2} \right)
		%e^{-(x^2 + y^2)/w_0^2}
	\label{eq:EffectiveA}
	\end{equation}
where $w_0$ is the laser waist.
The number of electrons that encounter the pulse peak at $(x,y)$, experiencing
a effective $a_0$ given by \cref{eq:EffectiveA}, is
	\begin{equation}
	\rmd N_e =
		\frac{N_{e,\mathrm{b}}}{\pi R^2}
		\exp\!\left( -\frac{(x - \Delta)^2 + y^2}{R^2} \right)
		\rmd x\,\rmd y.
	\end{equation}
The total number of positrons produced by a beam $N_{+,\mathrm{b}} = \int\!N_+(x,y)\,\rmd N_e$
where $N_+(x,y)$ is obtained by replacing the $a_0$ in \cref{eq:PositronYieldE}
with $a$ as given by \cref{eq:EffectiveA}.

We take as an example the collision between a $2\,$GeV electron beam
(total charge $100\,$pC, spherically symmetric FWHM $10\,\micron$, $R = 6\,\micron$) and
laser pulse with peak intensity $5\times10^{21}\,\Wcm^{-2}$, wavelength
$0.8\,\micron$, FWHM $30\,\fs$ and waist $2\,\micron$ and
compare our predictions to the result of full-scale 3D PIC simulation (see
\cref{app:pic} for details).
Were we to take this as a plane-wave interaction, we would predict a positron yield
of $6.9\times10^{5}$ using \cref{eq:PositronYieldE}.
Accounting for beams' finite sizes by integrating $\int\!N_+(x,y)\,\rmd N_e$
numerically using \cref{eq:PositronYieldE,eq:EffectiveA}, we find $N_{+,b} \simeq 10500$.
This is in good agreement $N_{+,b} \simeq 10000$ from simulation and should be
assessed in light of the non-perturbative growth in the pair
creation probability shown in \cref{fig:PositronTotal}.
We find also that final energy of the positron beam predicted by
\cref{eq:PositronEnergy,eq:OmegaC}, $320\,\MeV$, is consistent with
the spectrum shown in \cref{fig:SimulationSpectra}.
With the addition of a perpendicular offset of $5\,\micron$ between the beams,
numerical evaluation predicts that the positron yield should be reduced to
$5300$; we find $N_{+,b} = 4700$ PIC simulation.

To provide an approximate analytical scaling for the number of positrons
produced by an electron beam of finite size, we evaluate
the integral $\int\!N_+(x,y)\,\rmd N_e$ with the Laplace method,
assuming that $R \gg w_0$ and that the dominant contribution is that of the
region near $x = y = 0$ where the field amplitude is largest.
The fastest dependence on $a$ (and therefore $x,y$) in \cref{eq:PositronYieldE}
is that of the auxiliary function $\aux$, so we keep all other
factors constant when evaluating the Hessian of $N_+(x,y)$. We find that
the number of positrons produced by a $R$-radius beam of $N_{e,\mathrm{b}}$ electrons
colliding with a laser pulse of waist $w_0$ with perpendicular
offset $\Delta$ to be
	\begin{equation}
	N_{+,\mathrm{b}} \simeq
		\frac{0.727 a_0 \omega_0 \omega_c}{m^2}
		\frac{w_0^2 e^{-\Delta^2/R^2}}{R^2}
		N_+
		N_{e,\mathrm{b}}
	\label{eq:BeamPositronYield}
	\end{equation}
for $w_0 < R$ and $N_+$ given by \cref{eq:PositronYieldE}. The leading factor may
be taken to be roughly 0.25, as to have substantial pair creation at all, $\omega_c$
must be sufficiently large that $2 a_0 \omega_0 \omega_c / m^2
\gtrsim 1$. For the collision parameters given above, \cref{eq:BeamPositronYield}
predicts $N_{+,b} \simeq 18600$ and $9300$ for an offset of $0$ and $5\,\micron$
respectively, accurate to within a factor of two.

\Cref{eq:BeamPositronYield} indicates that the accuracy of alignment
between electron beam and laser pulse must be about
the size of the electron beam itself. It
suggests further that it is advantageous to focus the laser pulse as tightly
as possible, increasing $a_0$ at the expense of $w_0$. The latter only
enters the scaling quadratically, whereas $N_+$ grows much faster with
$a_0$ through its dependence on $\aux$ (\cref{eq:K}).
Analytical work on the effect of tight focussing has already
begun~\cite{DiPiazza,DiPiazza2}, going beyond the plane-wave approximation
to explore the effect of wavefront curvature on the positron yield.
Nevertheless, as near-term experiments are likely to focus the intense
laser with optics with $f$-number closer to 2, the effect of finite
size and alignment errors are more significant.
%A full analysis of the positron yield in laser-electron beam collisions,
%accounting for the generation of gamma rays, finite sizes and offsets as
%we have done here, but in the tightly-focussed case is beyond the scope of this work.

\section{Summary}

The collision of an intense laser pulse with a high-energy electron beam 
is a promising experimental geometry for the production of
high-energy photons and positrons. We have presented analytical
expressions for the electron beam's energy loss, quantum non-linearity parameter
and self-consistent emission spectrum.
Our scaling law for the number of positrons produced by the emitted gamma rays
demonstrates good agreement with full-scale PIC simulation even when
the finite sizes of the colliding beams are accounted for.
We have shown that a near-term
experiment employing the collision of a $2\,$GeV electron beam and laser pulse
of intensity $5\times10^{21}\,\Wcm^{-2}$ will produce a
positron beam with energy $300\,\MeV$ and particle number $10^4$.
Experimental detection of this beam will provide unambiguous evidence
of pair creation via the non-linear Breit-Wheeler process.

\begin{acknowledgments}
The authors acknowledge support from the Knut and Alice
Wallenberg Foundation (T.B., M.M.), the Swedish Research Council
(grants 2012-5644 and 2013-4248, M.M.) and the European Union's
Horizon 2020 research and innovation programme under the Marie
Sk{\l}odowska-Curie grant No.~701676 (A.I.).
Simulations were performed on resources provided by the Swedish National
Infrastructure for Computing (SNIC) at the High Performance Computing
Centre North (HPC2N).
\end{acknowledgments}

\appendix
\section{Pair creation rate}
\label{app:pcr}

The probability rate of pair creation $W_\pm$ for a photon with energy $\omega$ and
quantum non-linearity parameter $\chi_\gamma = e \sqrt{-(F\cdot k)^2} / m^3$ is conveniently expressed as
	\begin{equation}
	W_\pm = \frac{\alpha m^2 \chi_\gamma T(\chi_\gamma)}{\omega}
	\end{equation}
using the auxiliary function~\cite{Erber,BKS}
	\begin{equation}
	T(\chi_\gamma) =
		\frac{1}{6\sqrt{3}\pi \chi_\gamma}
		\int_1^\infty \!
			\frac{8u+1}{u^{3/2}\sqrt{u-1}}
			K_{2/3} \!\left( \frac{8u}{3\chi_\gamma} \right)
		\rmd u.
	\end{equation}
(The integrand cannot be interpreted as a spectrum.) It has limits
	\begin{equation}
	T(\chi_\gamma) \simeq
		\begin{cases}
		\frac{3\sqrt{3}}{16\sqrt{2}} \exp\!\left( -\frac{8}{3\chi_\gamma} \right) & \chi_\gamma \ll 1, \\
		\frac{0.37961}{\chi_\gamma^{1/3}} & \chi_\gamma \gg 1
		\end{cases}
	\end{equation}
and the following approximation from \citet{Erber} works well across the
full range of $\chi_\gamma$:
	\begin{equation}
	T(\chi_\gamma) \simeq
		\frac{0.16}{\chi_\gamma}
		K_{1/3}^2 \!\left( \frac{4}{3\chi_\gamma} \right).
	\end{equation}

\section{Simulations}
\label{app:pic}

In \cref{sec:grp,ssec:ElectronBeam} we compare our theoretical prediction with
the results of single-particle Monte-Carlo simulations, using the same
code developed for and described in \cite{Blackburn,Blackburn2}.
Each electron is followed along its trajectory through the laser pulse
and QED events are sampled at every timestep using the standard Monte Carlo
approach~\cite{Duclous}.
The electron momentum is updated at every timestep assuming that the
external fields are constant and crossed, but the particle push
is simplified to ballistic propagation at the speed of light.
This requires $\gamma \gg a_0$ and restricts use of the code to
collisions with externally imposed electromagnetic
waves, but these approximations permit substantial speedup over
conventional PIC codes.

	\begin{figure}
	\includegraphics[width=0.8\linewidth]{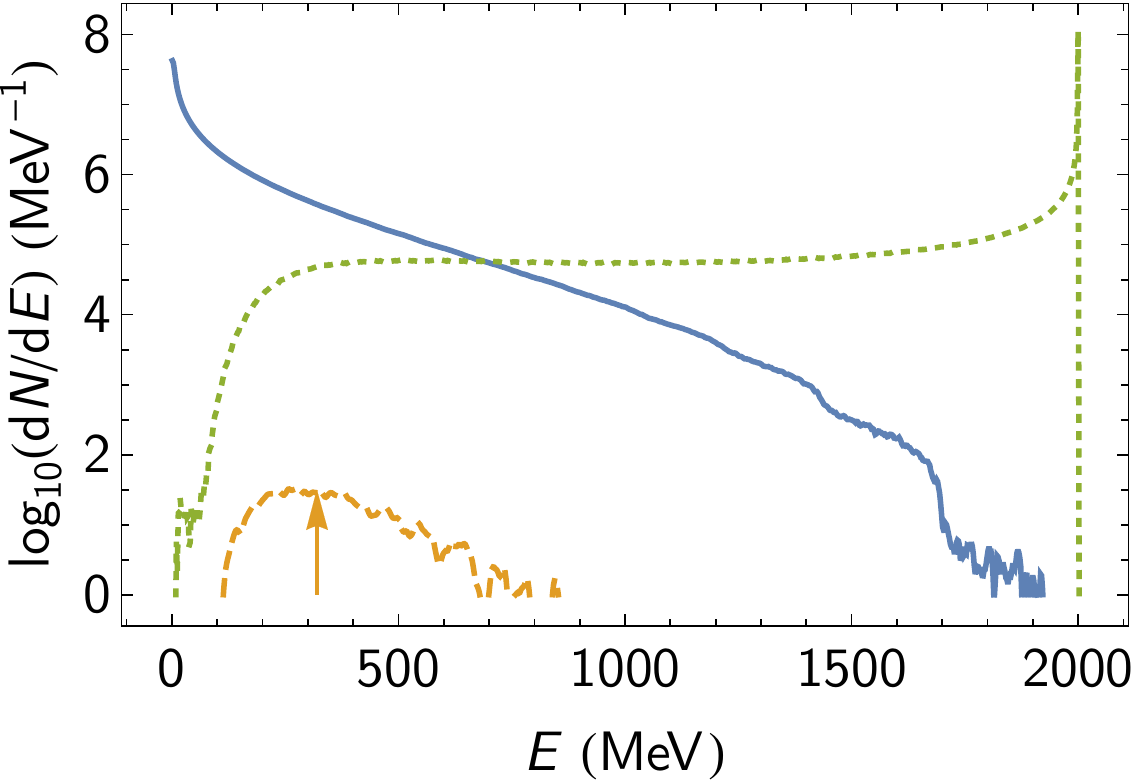}
	\caption[Simulation spectra]
			{The energy spectra of (blue, solid) photons, (yellow, dashed) positrons and
			(green, dotted) beam electrons
			emerging from a collision between an electron
			beam with energy $2\,\text{GeV}$ and a laser pulse with peak intensity
			$5\times10^{21}\,\Wcm^{-2}$ (see \cref{ssec:Expt} for details).
			(yellow arrow) The positron energy predicted by
			\cref{eq:PositronEnergy,eq:OmegaC}.}
	\label{fig:SimulationSpectra}
	\end{figure}

In \cref{ssec:Expt} we compare our theoretical predictions to the results of
simulations performed with the 3D PIC code \textsc{EPOCH}~\cite{Epoch}.
The first-order QED processes of photon emission and pair creation
are implemented via Monte-Carlo sampling of rates calculated in the
locally-constant-field approximation.
The simulation domain is $-10 < x/\micron < 10$,
$-10 < y/\micron < 10$ and $-10 < z/\micron < 30$, resolved with $(20,20,40)$
cells per micron in the $x$-, $y$- and $z$-directions respectively.
The timestep is set by the Courant-Friedrichs-Lewy condition, as there are sufficient cells
to ensure that the probability of multiple QED events in
a single timestep is negligible~\cite{RidgersJCP}.
The electron beam is initialized with Gaussian charge density profile
(FWHM $10\micron$) centred at $x = \Delta$, $y=0$ and $z = 24\,\micron$ (where
$\Delta$ is an offset between the beams), and represented with
8 macroelectrons per cell for a total of $9.9\times10^8$ particles.
The laser pulse is represented by a paraxial
Gaussian beam (waist $2\,\micron$, wavelength $0.8\,\micron$), is polarized
along $x$, propagates towards positive $z$ with
Gaussian temporal profile (intensity FWHM $30\,\fs$), and is timed to reach
focus when the electron beam centre arrives at $z=0$.
Final energy spectra for the collision parameters given in
\cref{ssec:Expt} are shown in \cref{fig:SimulationSpectra}.

\bibliography{references}

\end{document}